\documentclass[usenatbib]{aa}           
\usepackage[varg]{txfonts}

\usepackage[T1]{fontenc}
\usepackage{ae,aecompl}

\usepackage{graphicx}	
\usepackage{amsmath}	
\usepackage{amssymb}	
\usepackage{array}

\def\mod#1{{#1}}
\def\modb#1{{#1}}

\def\figsiz{8.5cm}

\newcommand\figa{
\begin{figure}[h]
\includegraphics[width=\figsiz]{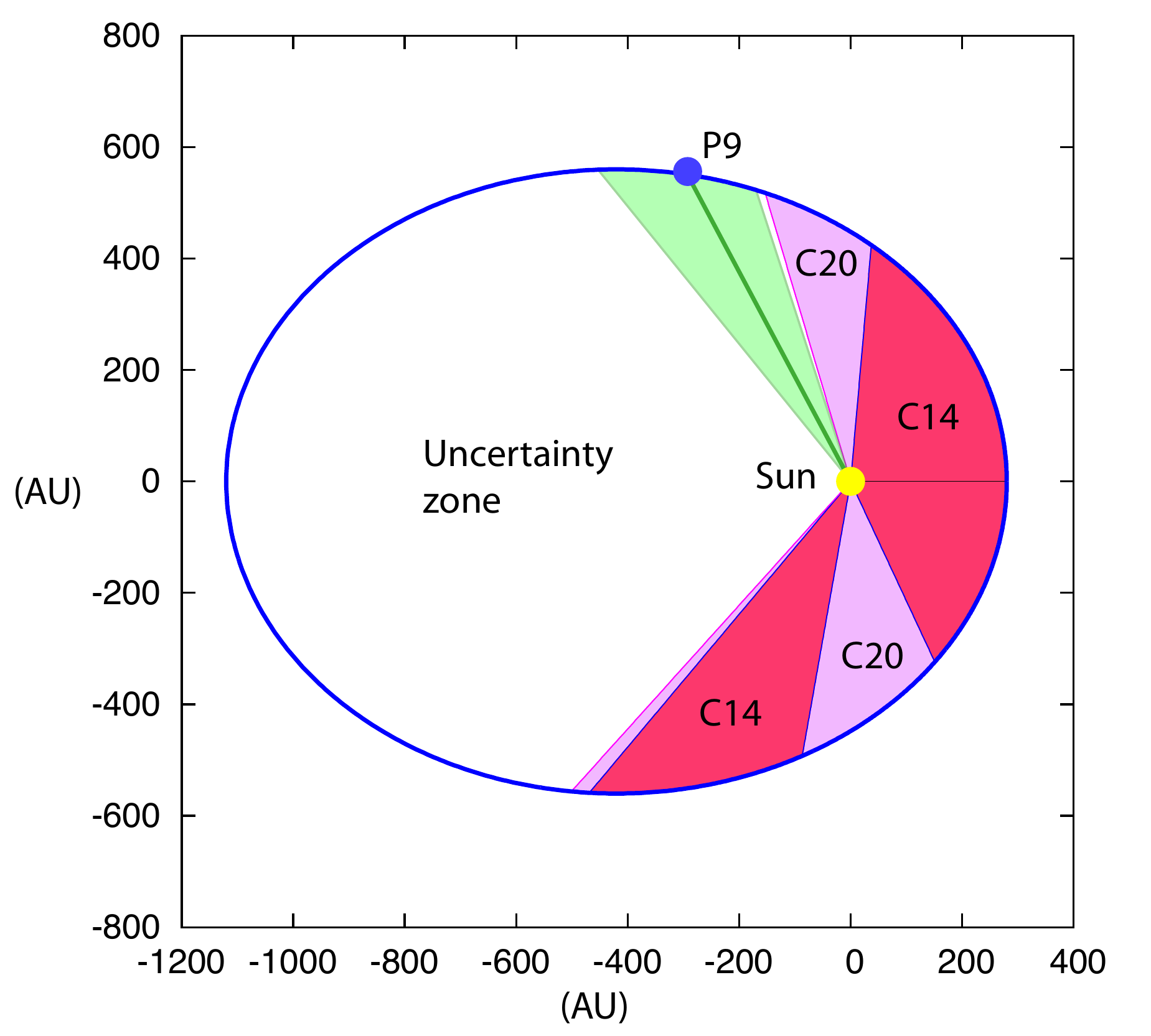}
\caption{Allowed zone for P9. The red zone (C14) is excluded by the analysis of the Cassini data up to 2014 (Sec.\ref{results}). 
The pink zone (C20) is how much this zone can be enlarged by extending the Cassini data to 2020 (Sec.\ref{extrap}).
The green zone is the most probable zone for P9 ($v\in [108^\circ:129^\circ]$),  with maximum reduction of the residuals 
 at $v=117.8^\circ$ (blue dot P9). The white zone is the uncertainty zone where the P9 perturbation is too faint to be detected.}
\label{secteur}
\end{figure}
}

\newcommand\figb{
\begin{figure}[h]
\includegraphics[width=\figsiz]{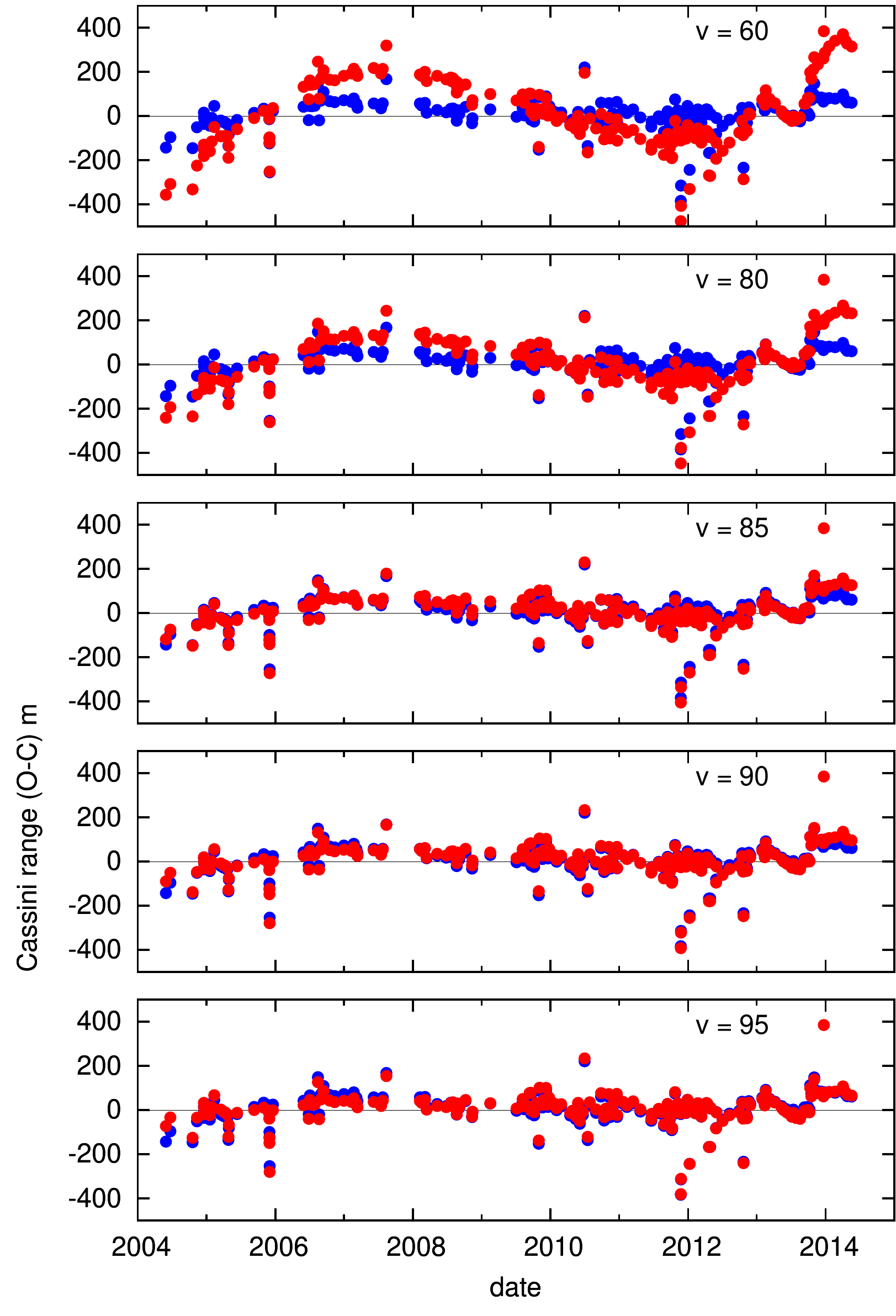}
\caption{Post fit residuals (in m) for the Cassini observations over $[2004.4:2014.4]$. In blue are the nominal residuals of INPOP. In red are the post fit residuals after the introduction of P9 in the INPOP model. }
\label{res_cas}
\end{figure}
}

\newcommand\figc{
\begin{figure}[h]
\includegraphics[width=\figsiz]{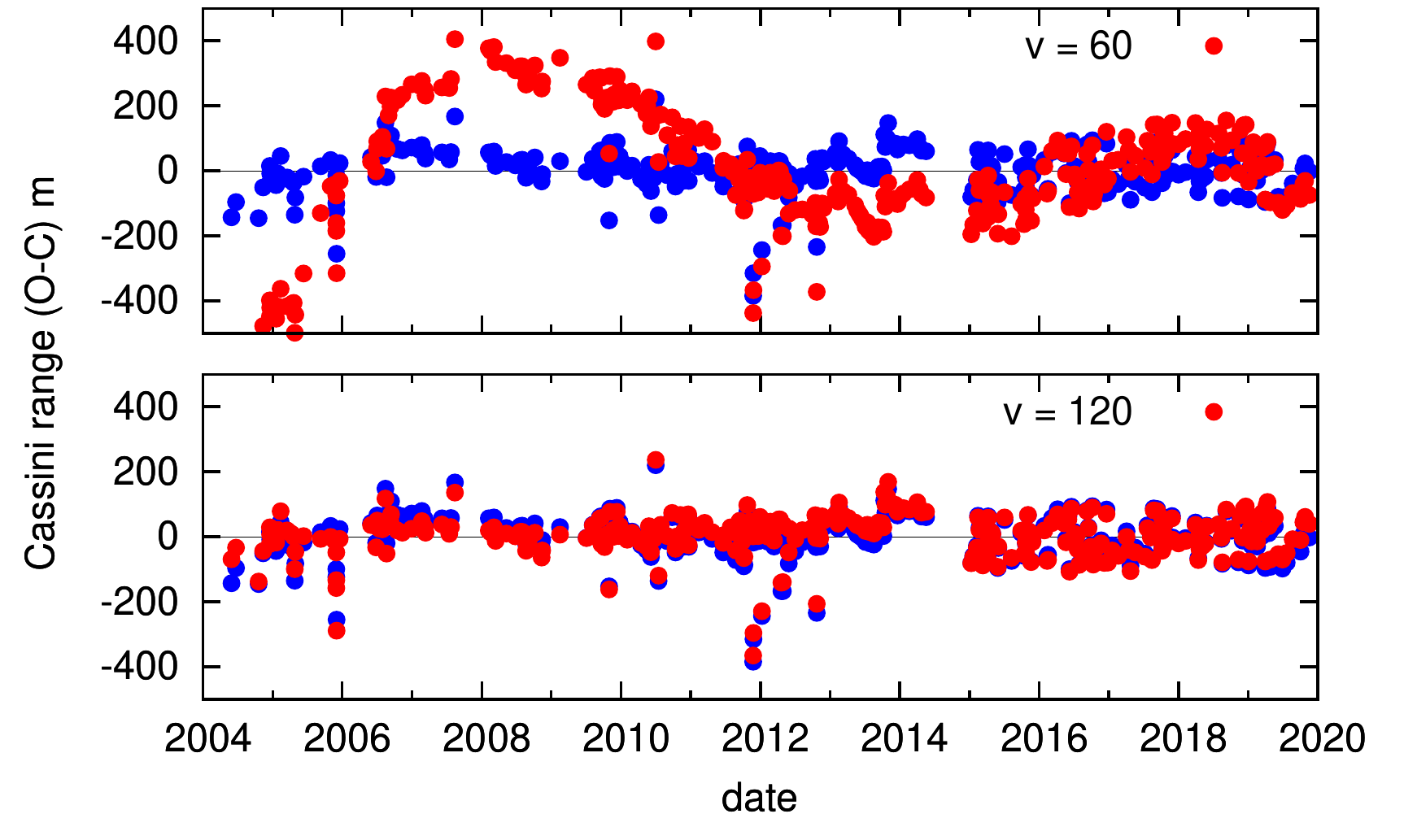}
\caption{\mod{Same as Fig.\ref{res_cas} with the addition of simulated Cassini data over $[2014.4:2020]$,
 for $v=60^\circ$ and  $v=120^\circ$.} }
\label{extrap}
\end{figure}
}

\newcommand\figd{
\begin{figure}[h]
\includegraphics[width=\figsiz]{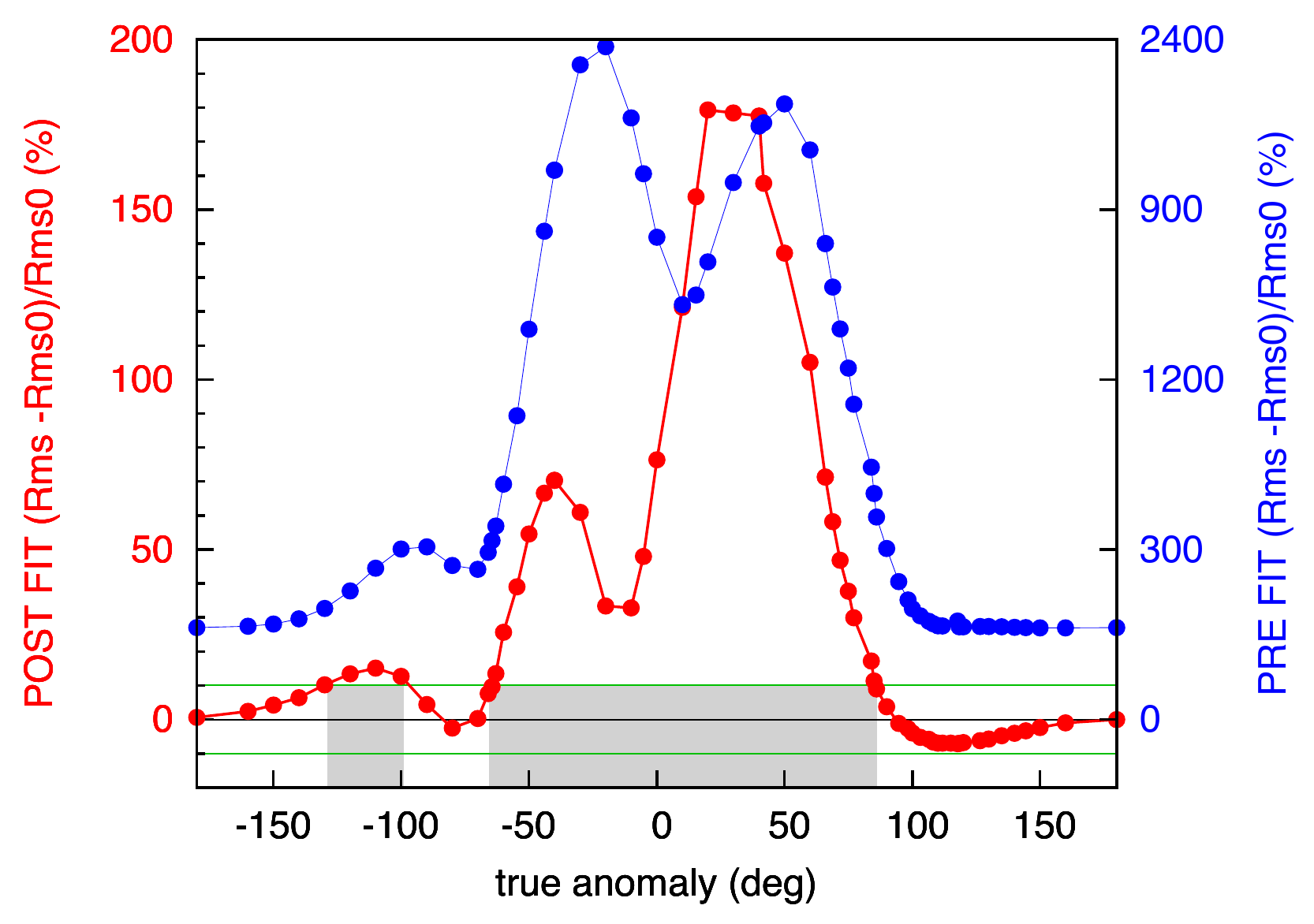}
\caption{Relative change in the root mean square  residuals  of a fitted INPOP solution including P9, with respect to those of the nominal INPOP,  with respect to the  true anomaly of P9 ($v$ in $x$ axis). Pre-fit  residuals are in blue with the right y-scale (in $\%$). 
Post-fit residuals are in red with the left y-scale (in $\%$).}
\label{rms}
\end{figure}
}

\newcommand\figdd{
\begin{figure}[h]
\includegraphics[width=\figsiz]{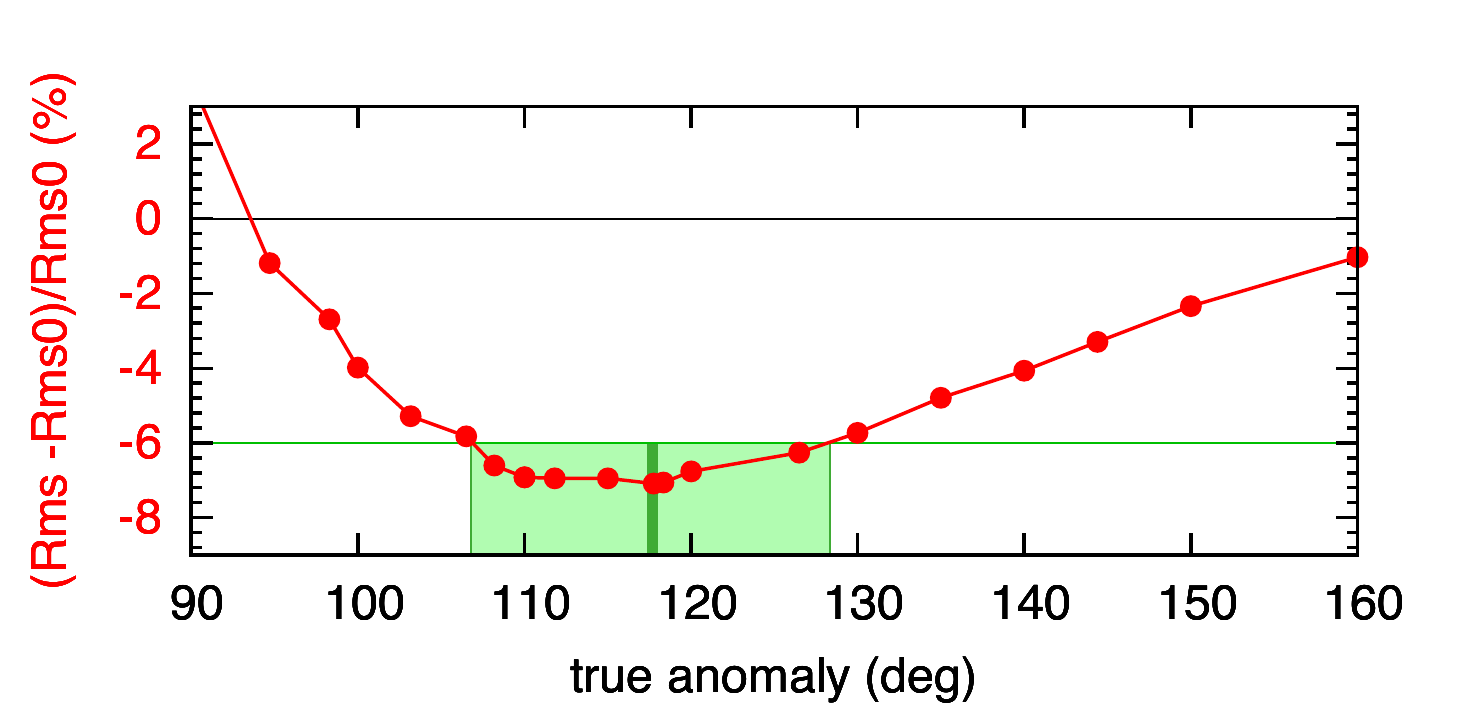}
\caption{\mod{Enlargement of Fig.\ref{rms} around the minimum (green bar,  at $v=117.8^\circ$). 
In the green zone, $\Delta\sigma < -6\%$.}}
\label{rmsE}
\end{figure}
}

\newcommand\fige{
\begin{figure}[h]
\includegraphics[width=\figsiz]{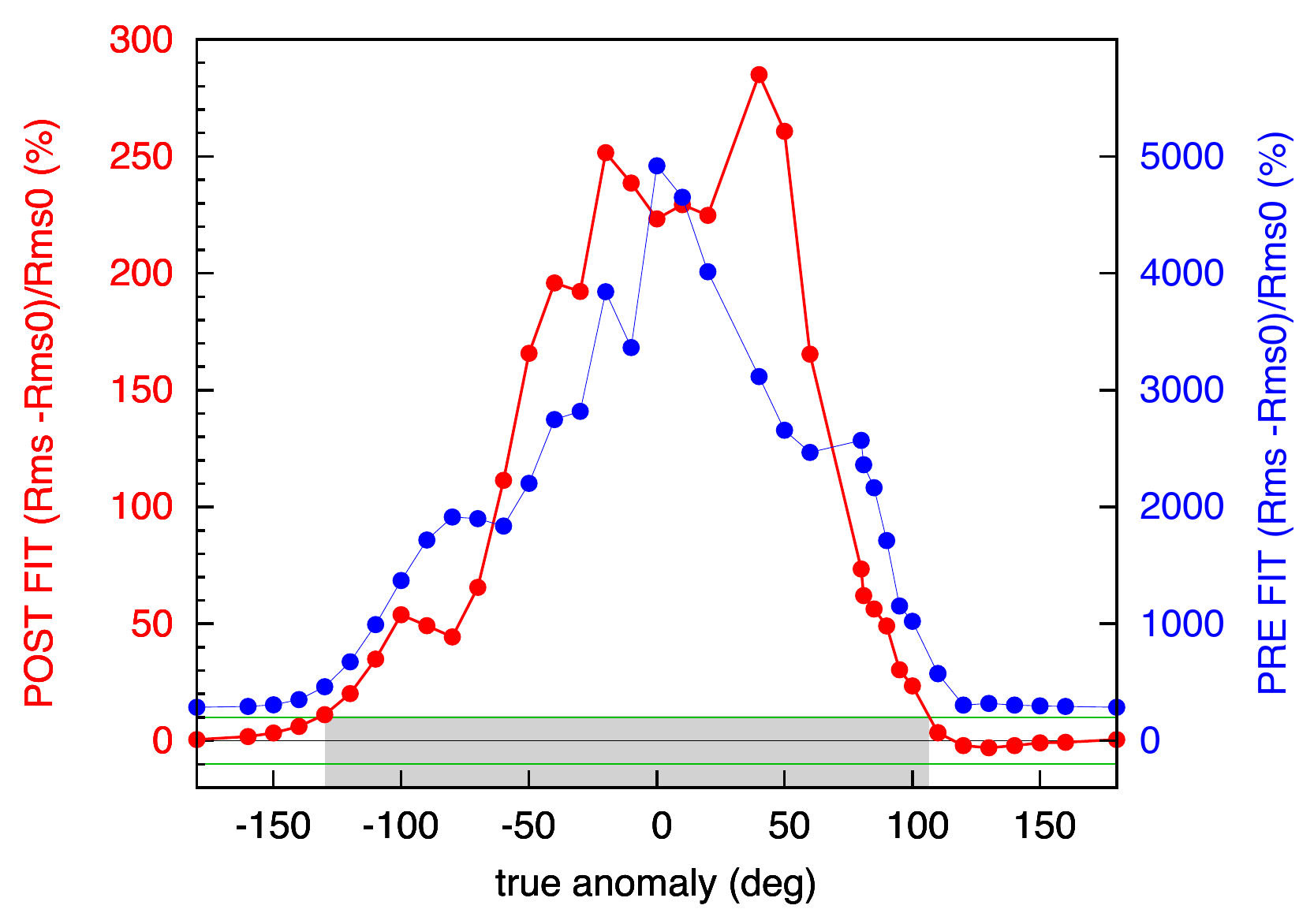}
\caption{\mod{Same as Fig.\ref{rms}  with extrapolated 
 Cassini data \modb{(wihout P9)} up to 2020.}}
\label{rmsx}
\end{figure}
}

\begin{document}

\title{Constraints on the location of a possible  9th planet derived from the Cassini data }

\author{
A. Fienga\inst{1,2}
\and J. Laskar\inst{2}
\and H. Manche\inst{2}
\and M. Gastineau\inst{2}
}

\titlerunning{Contraints on planet 9}
\authorrunning{Fienga et al.}

\institute{
G\'eoAzur, CNRS-UMR7329, Observatoire de la C\^ote d'Azur, Universit\'e Nice Sophia Antipolis, 250 Av. A. Einstein, Valbonne, 06560, France\\
\email{fienga@oca.eu}
\and ASD/IMCCE, CNRS-UMR8028, Observatoire de Paris,  PSL, UPMC, 77 Avenue Denfert-Rochereau, 75014 Paris, France\\
\email{laskar@imcce.fr}
}

\date{Accepted 16 February 2016. Received 31 January 2016.}


\abstract{
To explain the unusual distribution of Kuiper Belt objects, several authors have advocated the existence of a super-Earth planet in the 
outer solar system. It has recently been proposed that a 10  M$_{\oplus}$ object with an orbit of 700 AU 
semi major axis and 0.6 eccentricity can explain the observed distribution of Kuiper Belt objects around Sedna. 
Here we use the INPOP planetary ephemerides model as a sensor for testing for an 
additional body in the solar system. We test the possibility of adding the proposed planet without 
increasing the residuals of the planetary ephemerides, fitted over the whole INPOP planetary data sample. 
We demonstrate that the presence of such an object is not compatible  with
the most sensitive data set, the Cassini radio ranging data, if its true anomaly is in the intervals $[-130^\circ:-100^\circ]$ or $[-65^\circ : 85^\circ]$.
Moreover, we find that the addition of this object can reduce the Cassini residuals, with a  most probable position 
\mod{given by a true anomaly $v = {117.8^\circ}^{  + 11^\circ}_{ - 10^\circ} $}.
}


\maketitle



\section{Introduction}

The discovery in 2014 of the Kuiper Belt object (KBO) 2012 VP$_{{113}}$ \citep{Trujillo14} 
in the inner Oort cloud revived the question of a planet X. 
With a relatively large radius  (R $\approx$ 200-1000km), this object has orbital parameters similar to those of Sedna, another large planet-like KBO, with a high eccentric orbit (e>0.7), a perihelia distance greater than 30 AU, and arguments of perihelia $\omega$ around 0$^{\circ}$.
As stressed in \citep{Trujillo14,FuenteMarcos2014}, the recent observed distribution of $\omega$ for the inner Oort cloud objects is concentrated around 0$^{\circ}$. 
One proposed explanation for such an unlikely spatial distribution  is the existence of a super-Earth mass planet with mass between 2 to 15 M$_{\oplus}$ and a semi-major axis of 200 to 300 AU \citep{Trujillo14}. 

The existence of such object can have an important impact not only  on the mass distribution of the material beyond the limit of 50 AU  but also on the spatial distribution of objects between 40 to 50 AU \citep{Lykawa2008,Kenyon15}. 
The present mass distribution in the outer edge of our solar system can be seen as a footprint of the mechanism of formation of our solar system. Characterization of the orbit parameters of a super-Earth beyond 200 AU can give some estimations of the mass and the outer radius of the solar nebula \citep{Bromley2014} as well as  some clues to the scenario of formation. Low eccentric orbits is compatible with migrating scenario \citep{Ward1997,Crida2009}, higher eccentricity for low mass planet being more compatible with a gas drag formation model \citep{Kenyon15}. 
Such an object  has also been invoked to explain some  periodicities in mass extinction events
\citep{raup1984,Whitmire2015}. 

The question of  the existence of a massive object  in the outer solar system thus becomes   important  for studying  the present solar system's architecture and understanding its formation. 
Very recently, this quest has reached a new level with the publication by \citet{BatyBrow2016a} of a proposal of a ninth planet to explain the clustering of the distant KBO  in the vicinity of 2012 VP$_{{113}}$. The gravitational interaction with this object 
(which we will call P9 from now on) should have scattered the bodies that are not confined in the vicinity of the central island of a  pseudo resonance
\footnote{We denote this kind of resonance a pseudo resonance, because it does not involve the existence of separatrices in the phase space \citep{BatyBrow2016a}}
 with P9. To be able to remove the bodies that were originally evenly distributed in a disk, P9 needs to be massive enough and not too far, with high eccentricity.   \citet{BatyBrow2016a} propose a $10\,$ M$_{\oplus}$ planet with 0.6 eccentricity and a 700 AU semi major axis. In the present work, we examine the contraints given by the analysis of the dynamics of the solar system on the possibility of such a new planet. 

\section{Searching for planet X}

The hunt for a planet X  started in 1915 \citet{Lowell1915}, and some important limitations were provided by different approaches.  The direct imaging survey of the far solar system by infrared space telescopes  IRAS and WIZE 
 did not detect any massive objects of  Jupiter size inside 25000 AU and of  Saturn size inside 10000 UA \citet{Luhman14}.
In 1993, \citet{Standish1993} demonstrated that no anomalous residuals can be seen in the 
residual of the outer planet orbits. However since this work, no direct confrontation 
has been performed between the perturbations induced by a planet X on the orbits of the main planets of our solar system and the best fitted planetary ephemerides.
The only estimates were made indirectly, based on the results of the ephemerides, but without refitting the whole parameters of these ephemerides  \citep{Iori2012a,Iori2016a}. 

Since 1993, very accurate observations of Saturn orbit were obtained thanks to the tracking of the Cassini spacecraft during its exploration of the Saturnian system. As described in \citep{2014IPNPR.196C...1F}, \citep{Hees2014a} and \citep{Fienga2016}, the ten year positions of Saturn allow significant improvement in our knowledge of Saturn's orbit, as well as of  Jupiter, Uranus, and Neptune orbits. These Cassini data have  already been used very successfully to 
put some strict limits on the possibility of an anomalous Pioneer acceleration \citep{AndeLain2002a,FienLask2010a}.
Furthermore, thanks to the Cassini flyby of Jupiter, a supplementary accurate position of Jupiter is also used to build the ephemerides.
Finally, the flyby of the New Horizons spacecraft of the Pluto-Charon system should bring some supplementary information and constraints for the existence of a super-Earth.

We use here the dynamical model of the INPOP planetary ephemerides \citep{Fienga2008,FienLask2009a,FienLask2010a,Fienga2011,Fienga2016} 
for testing the possibility of an additional planet, focussing on the proposed nominal planet P9 of  \citet{BatyBrow2016a}.
In the dynamical model of INPOP  planetary ephemerides, we add  a super-Earth object of $10\,$ M$_{\oplus}$  with different orbital characteristics, in agreement with the proposed orbit of P9. We then build  new ephemerides including these objects and perform a global fit of the perturbed planet orbits to the whole data set that is used to construct the INPOP and DE430 JPL ephemerides  \citep{2014IPNPR.196C...1F}.

\figb

\section{Method}
\label{sec1}

\subsection{INPOP planetary ephemerides}

Since \citep{Fienga2008}, the INPOP planetary ephemerides are regularly produced and used  for testing alternative theories of gravity,  studying variations in the solar plasma density, and estimating asteroid masses \citep{FienLask2009a,FienLask2010a,Fienga2011,Fienga2016}.
The INPOP ephemerides of the eight planets of our solar system, of Pluto, and  of the Moon, are obtained by numerically integrating  the barycentric 
Einstein-Infeld-Hoffmann equations of motion \citep{Moyer1971} in a suitable relativistic time-scale, and taking up to 298 asteroids of the main belt into account. 

In 2015, we released the latest INPOP ephemerides, built over the whole sample of modern and ancient planetary observations, from ten years of Cassini round-trip around Saturn and its system, up to the first photographic plates of Pluto. Besides the differences in the dynamical modeling and in fitting choices explained in \citet{Fienga2016}, this  latest version of INPOP is very close to the JPL DE430 planetary ephemerides \citep{2014IPNPR.196C...1F}.

\subsection{INPOP: a tool to test the presence of P9 }

To test the possible presence of P9, we add it to the dynamical model of the solar system in INPOP. 
We then proceed to a complete fit of all the initial conditions and parameters for the planetary orbits of INPOP, 
totalling 150  variable quantities. The fit is performed over the full set of  about 150 000 planetary and satellite 
observations that are currently used 
to determine the INPOP ephemerides, although the INPOP ephemerides also comprise  
the adjustment of the parameters of the Moon which will not be considered here. 
The presence of P9 is only possible if the post-fit  residuals are not increased after the introduction of P9 in the INPOP model. 
The increase of the residual will be a clue that P9 cannot exist at the given location. 
In the same way, a significant decrease in the residuals could be the signature of P9.

\begin{table}
\caption{Orbital elements of P9, as given in \citet{BatyBrow2016a}, in ecliptic (IERS) orbital elements 
$(a,e,i,\omega,\Omega)$, respectively semi-major axis, eccentricity, inclination, argument of perihelion, and longitude of the node. }
      \begin{tabular}{ccccc}
\hline\noalign{\smallskip}
$a$ (AU)  & $e$  & $ i$ (deg) & $\omega$ (deg) & $\Omega$ (deg) \\ 
\noalign{\smallskip}
700  & 0.6 & 30 & 150 & 113 \\
\hline
      \end{tabular}
      \label{param}
\end{table}

For P9, we use the ecliptic orbital parameters of \citet{BatyBrow2016a} (Table \ref{param}). 
In \citet{BatyBrow2016a}, only the long term evolution is considered as a constraint for the distribution of the KBO, so the true anomaly is left as an unknown. With a large eccentricity of $e=0.6$, the radius vector from the Sun, $r$, can vary from 
280   to  1120 AU, depending on the true anomaly $v$ of P9. We  thus sample  the possible values of $v$   over a full orbit ($ v \in [-180^\circ:180^\circ]$).

\section{Results}
\label{results}
\figd
\mod{\figdd}

In figure \ref{res_cas} we represented the post fit residuals of the INPOP ephemerides after the addition of P9 for various values of the true anomaly $v = 60^\circ, 80^\circ, 85^\circ, 90^\circ, 95^\circ$. For comparison, the INPOP residuals are also plotted. 
We have not plotted 
the residuals for values of $v$ less than $60^\circ$, as for $v=60^\circ$, it is already obvious that P9 cannot exist. 
For $v=95^\circ$ the difference  in the residuals is too small to be significant, while these residuals already increase for $v=85^\circ$.
In figure \ref{rms}, we have gathered  the results computed for all values of $v$ by plotting,
for the pre-fit ($\tilde\Delta\sigma$, in blue) and post-fit ($\Delta\sigma$, in red) residuals, the 
relative change in root mean square  residuals ($\Delta\sigma  = (\sigma-\sigma_0)/\sigma_0$), where 
$\sigma_0=74.9$m is the standard deviation for the INPOP nominal ephemerides over the 
Cassini data.
 
The pre-fit residuals are differences between the observations 
(here, radio ranging) and the computed values of observed quantities, for INPOP ephemerides, with the addition 
of P9 to the INPOP model, without fitting the parameters.
The variation in these residuals  with the  true anomaly reflects how much the perturbation  of P9 on 
Saturn changes with the geometry of the problem, and not only with the distance of P9 from the Sun. 
The post-fit $\sigma$ 
are the residuals after the fit of 56 planetary parameters in the INPOP ephemerides after the addition of P9 in the model. 
This fit is performed over the whole set of more than 150000 planetary observations. 

The shaded region corresponds to a relative increase in the post-fit residuals of more  than $10\%$  
after the  addition  of P9. This   is about the level of  \modb{precision
of  the INPOP $\sigma_0$, estimated by comparison with DE430.} We thus consider that an increase of  $\Delta\sigma$ above  $10\%$  
is significant and denotes the  impossibility of fitting a  P9 planet for this value of the 
true anomaly $v$. The shaded regions ($[-130^\circ:-100^\circ] \cup [-65^\circ:85^\circ]$) are thus the forbidden regions for P9.
The $\Delta\sigma$  curve presents two minima. We do not believe that the minimum at $v=-80^\circ$ 
is really significant, since it also corresponds to a minimum of the pre-fit residuals and amounts to only $\Delta\sigma=-2.5\%$.

The other minimum  at  $v=117.8^\circ$ is more interesting. It corresponds both to the observed 
minimum of the residuals curve (Fig.\ref{rmsE}), with 
$\Delta\sigma= - 7.09\%$, and  to the result of a direct fit of the mean anomaly over the Saturn Cassini range data sample
that provided an uncertainty at 2-$\sigma$ of $\pm 1.8^\circ $ on $v$.
Owing to the relatively flat behaviour  around the minimum in Fig.\ref{rmsE}, we prefer  a conservative approach here, 
and we defined 
an empirical uncertainty  by the zone (in green in Fig.\ref{rmsE}) for which $\Delta\sigma < -6\%$, which 
gives\footnote{At this point, no false alarm probability has been associated to this range of values.} 
$v={117.8^\circ}^{  + 11^\circ}_{ - 10^\circ}$.

\section{Extrapolation of Cassini data}
\label{extrap}
Although the INPOP ephemerides were fitted to all 150 000 planetary observations, 
we have shown here only the residuals with the Cassini data because they are
 most sensitive to the presence of an additional perturber P9. 
We have only used the data available up to 2014.4, but 
new data will be available until the end of the mission, which is programmed for 2017, unless the mission is extended.
To evaluate the impact of an increase in the time span  of the Cassini mission, 
we have simulated additional data, by adding to the INPOP prediction (without P9) a  Gaussian noise of 200m sigma, 
which is slightly pessimistic with respect to the  already analysed data set. 
As previously, P9 is then added, and the ephemerides are fitted to all observations, including the extrapolated Cassini data,
 through  several iterations (Figs.\ref{extrap}, \ref{rmsx}). 
Extending Cassini mission until 2020  would thus allow to state for the non existence of P9 on the 
interval  $v\in [-132^\circ: 106.5^\circ]$ (Fig.\ref{rmsx}). 
 The respective area  of exclusion of a possible P9 from the present data and 
from the extrapolated  Cassini data up to 2020 are displayed in Fig.\ref{secteur}, together with the most probable 
zone for finding P9.
\figc
\fige


\section{Conclusions}
\figa
The Cassini data provides an exceptional set of measures that acts as a very sensitive device 
for testing the possibility of an additional massive body in the solar system. With the data accumulated until 2014.4, 
we can exclude the possibility  that P9 is in the section of the orbit 
depicted in red in Fig.\ref{secteur}, with a true anomaly $v$ in $ [-130^\circ:-100^\circ] \cup [-65^\circ:85^\circ]$.
We thus contradict the affirmation of \citet{Iori2016a},  who states that a body of 10 M$_{\oplus}$ 
is excluded if it resides closer to 1000 AU of the Sun. \citet{Iori2016a} does not properly consider 
how much the presence of an additional body can be absorbed by the fit of all the other parameters in the solar system ephemerides. 
The global fit that we present here avoids this drawback. 
Moreover, we found that the presence of P9 could significantly  decrease the Cassini residuals  if $v$ is in 
the interval $[108^\circ:129^\circ] $, 
with a most probable position at $v={117.8^\circ}^{  + 11^\circ}_{ - 10^\circ} $.

Since the Cassini data is at present the most precise sensor for testing the possibility of an additional body in the solar system, it is essential that Cassini continues to provide ranging data, since there will not be very soon an additional spacecraft around one of the planets beyond Saturn. 
Extending the Cassini data up to 2020 will already allow to state for the existence of P9 for  $v \in [-132^\circ: 106.5^\circ]$.
Juno will soon arrive around Jupiter and will thus allow us to improve the orbit of Jupiter. 
This may not directly improve the constraints on P9, because Jupiter is less sensitive than Saturn to the perturbations of P9.
 Nevertheless,  constraining Jupiter  more  tightly  will allow us to improve the determination of P9, 
because less flexibility will exist for absorbing the perturbations due to P9. 




\bibliographystyle{aa}
\bibliography{biblio_PX} 


\label{lastpage}
\end{document}